\documentclass[aps,showpacs,prb,twocolumn,superscriptaddress,floatfix,longbibliography]{revtex4-2}

\usepackage{graphicx}         
\usepackage{epstopdf}         
\usepackage[colorlinks=true,linkcolor=blue,citecolor=blue]{hyperref} 
\usepackage{url}              

\usepackage{amsmath, amssymb} 
\usepackage{mathtools}        
\usepackage{bm}               
\usepackage{bbm}              
\usepackage{amsthm}           

\usepackage{float}            
\usepackage[section]{placeins}
\usepackage{booktabs}         
\usepackage{tabularx}         
\usepackage{multirow}         
\usepackage{threeparttable}   
\usepackage{makecell}         
\usepackage{diagbox}          
\usepackage{array}            
\usepackage{adjustbox}        
\usepackage{enumitem}         

\usepackage{algorithm}
\usepackage{algpseudocode}    
\usepackage{tikz}             
\usepackage{empheq}           

\usepackage{xspace}           
\usepackage{color}            
\usepackage{siunitx}          

\begin{document}
\title{Universality Diagram of Phase Transitions in Long-range Statistical Systems}

\author{Tianning Xiao}
\affiliation{Hefei National Research Center for Physical Sciences at the Microscale and School of Physical Sciences, University of Science and Technology of China, Hefei 230026, China}

\author{Zhijie Fan}
\email{zfanac@ustc.edu.cn}
\affiliation{Hefei National Research Center for Physical Sciences at the Microscale and School of Physical Sciences, University of Science and Technology of China, Hefei 230026, China}
\affiliation{Hefei National Laboratory, University of Science and Technology of China, Hefei 230088, China}
\affiliation{Shanghai Research Center for Quantum Science and CAS Center for Excellence in Quantum Information and Quantum Physics, University of Science and Technology of China, Shanghai 201315, China}

\author{Youjin Deng}
\email{yjdeng@ustc.edu.cn}
\affiliation{Hefei National Research Center for Physical Sciences at the Microscale and School of Physical Sciences, University of Science and Technology of China, Hefei 230026, China}
\affiliation{Hefei National Laboratory, University of Science and Technology of China, Hefei 230088, China}
\affiliation{Shanghai Research Center for Quantum Science and CAS Center for Excellence in Quantum Information and Quantum Physics, University of Science and Technology of China, Shanghai 201315, China}

\begin{abstract}
The percolation, Ising, and O($n$) models constitute fundamental systems in statistical and condensed matter physics. For short-range-interacting cases, the nature of their phase transitions is well established by renormalization-group theory. However, the universality of the transitions in these models remains elusive when algebraically decaying long-range interactions $\sim 1/r^{d+\sigma}$ are introduced, where $d$ is the dimensionality and $\sigma$ is the decay exponent. Building upon insights from L\'evy flight, i.e., long-range simple random walk, we propose three universality diagrams in the $(d,\sigma)$ plane for the percolation model, the O($n$) model, and the Fortuin-Kasteleyn Ising model, respectively. The conjectured universality diagrams are consistent with recent high-precision numerical studies and rigorous mathematical results, offering a unified perspective on critical phenomena in systems with long-range interactions.
\end{abstract}

\maketitle

\section{Introduction}
\label{sec:intro}

Universality lies at the heart of the renormalization-group (RG) theory of critical phenomena. Within this framework, the upper critical dimension, $d_c$, serves as a fundamental boundary separating distinct universality classes~\cite{PhysRevB.4.3174, PhysRevB.4.3184, PhysRevLett.28.240, chayes1987upper}. For dimensions $d < d_c$, critical behavior is governed by nontrivial RG fixed points, resulting in anomalous dimensions and nontrivial critical exponents. Conversely, for $d > d_c$, interactions become irrelevant, and the system falls into the mean-field (MF) universality class characterized by Gaussian MF exponents. At the upper critical dimension $d_c$, mean-field scaling laws are preserved, but modified by logarithmic corrections. Standard examples include O($n$) spin systems ($\phi^4$ theory) with $d_c=4$~\cite{PhysRevLett.28.240} and percolation ($\phi^3$ theory) with $d_c=6$~\cite{chayes1987upper}. Despite this well-established picture, the manifestation of universality in finite-size scaling (FSS) above $d_c$ remains subtle and controversial. Recent numerical studies reveal that, under periodic boundary conditions (PBCs), high-dimensional systems exhibit a characteristic two-scale FSS behavior~\cite{huang2018, fang_complete_2020, fang_logarithmic_2021, lv2021, PhysRevE.108.024129, PhysRevE.109.034125, PhysRevE.110.044140}. This phenomenon arises from the interplay between Gaussian fixed-point (GFP) scaling and complete graph (CG) scaling, where the latter corresponds to the Landau mean-field theory in a finite system. Specifically, macroscopic observables like magnetization and susceptibility are governed by CG exponents $(y_t^{\rm CG}, y_h^{\rm CG})$, whereas distance-dependent quantities and correlation functions are controlled by GFP exponents $(y_t^{\rm GF}, y_h^{\rm GF})$. This two-scale framework provides a crucial key to resolving long-standing debates regarding the critical scaling in high dimensions.

The high-dimensional critical behavior of O$(n)$ spin models and percolation can be intuitively understood through a geometric perspective from the simple random walk (SRW)~\cite{fernandez_random_1992, zhou_random-length_2018, hara_mean-field_1990}. For O$(n)$ models, the geometric mechanism depends on the spin component $n$. In the $n \to 0$ limit, the model corresponds to the self-avoiding walk (SAW). Above $d_c=4$, the self-avoidance constraint becomes RG-irrelevant, and the critical two-point function recovers the SRW form $G(r) \sim |r|^{2-d}$. For O$(n)$ models with $n\ge1$, the spin-spin correlation function maps to path weights with effective interactions determined by the intersection of trajectories~\cite{fernandez_random_1992}. Since Brownian paths have fractal dimension $2$, the intersection of two independent paths becomes marginal at $d_c=4$; above this, paths become mutually transparent, yielding Gaussian mean-field exponents. By analogy, high-dimensional percolation is governed by the intersection of three independent walks, which shifts the critical dimension to $d_c=6$~\cite{hara_mean-field_1990}. Consequently, the SRW provides the universal geometric backbone for the mean-field physics in these systems.

Complementing the thermodynamic perspective, the geometric representation of statistical systems reveals rich critical properties and offers profound insights into the original models. This is best exemplified by the Fortuin–Kasteleyn (FK) random-cluster representation~\cite{grimmett2006}. In this framework, the Ising model (equivalent to the $q=2$ Potts model) is reformulated in terms of bond configurations $A\subseteq E$ with the partition function:
\begin{align}
    \mathcal{Z}_{\rm FK} = \sum_{A\subseteq E} q^{k(A)} v^{|A|},
    \label{eq:FK}
\end{align}
where $k(A)$ is the number of clusters, $v$ is the bond weight, and $|A|$ is the number of occupied bonds. Ordinary percolation is recovered as $q\to 1$, while the Ising model corresponds to $q=2$. Recent numerical and mathematical advances have uncovered that the FK–Ising model possesses \emph{two} distinct upper critical dimensions~\cite{fang_geometric_2022, fang_geometric_2023, wiese_two_2024, engelenburg_one-arm_2025}. The thermodynamic upper critical dimension, $d_c=4$, is inherited from the $\phi^4$ field theory and controls standard thermodynamic observables. Meanwhile, FK clusters exhibit an additional geometric upper critical dimension, $d_p=6$. For $4 < d < 6$, the largest and second-largest clusters scale with distinct mean-field exponents derived from the CG and the GFP, respectively. In contrast, for $d \ge 6$, the FK clusters, except for the largest one, become percolation-like, characterized by strongly winding clusters and a diverging number of spanning clusters, consistent with the high-dimensional percolation model~\cite{fang_geometric_2022, fang_geometric_2023}. This double upper critical structure underscores the necessity of geometric observables for a comprehensive characterization of criticality.

While spatial dimension dictates the critical behavior in short-range (SR) models, the introduction of long-range (LR) interactions provides an alternative mechanism to tune universality, adding a new layer of complexity to the landscape of critical phenomena.
Since Dyson's seminal work on the one-dimensional Ising model with algebraically decaying couplings~\cite{dysonExistencePhasetransitionOnedimensional1969}, LR systems have been studied intensively~\cite{fisher1972,sak1973,luijtenBoundaryLongRangeShortRange2002, picco2012, angelini2014,defenu2023}. 
For a $d$-dimensional O($n$) spin model with algebraically decaying coupling strength, $J(r)\sim r^{-d-\sigma}$, the interplay between SR and LR terms generates rich critical phenomena in the $(d,\sigma)$ plane. 
Dimensional analysis of the effective Hamiltonian posits a threshold $\sigma_* = 2$ separating SR and LR universality, below which the LR interaction dominates, and the system enters the nonclassical regime where critical exponents are $\sigma$-dependent. For the O($n$) spin model, this nonclassical regime extends down to $\sigma = d/2$, below which the transition is governed by the long-range Gaussian fixed point (LR-GFP), and the system is in the MF regime. The $\epsilon$ expansion analysis by Fisher suggests that the anomalous dimension follows the LR-GFP prediction, $\eta = 2-\sigma$ up to $O(\epsilon^3)$ throughout the nonclassical regime~\cite{fisher1972}. Subsequently, Sak predicted that the SR universality remains stable down to $\sigma_* = 2 - \eta_{\rm SR}$, where $\eta_{\rm SR}$ is the anomalous dimension of the corresponding SR model~\cite{sak1973}. However, recent large-scale simulations of LR-O($n$) spin models, and LR-percolation models in two dimensions have challenged this picture, providing strong evidence that the SR-LR crossover is at $\sigma_*=2$~\cite{grassberger2013, LRIsing,liu_two-dimensional_2025,xiao_two-dimensional_2024,yao2025nonclassicalregimetwodimensionallongrange, LRHeisenberg}. These studies suggest that while the anomalous dimension $\eta$ follows the mean-field prediction $\eta=2-\sigma$ up to an extended range in the nonclassical regime, it exhibits nontrivial deviations as $\sigma \to 2^-$. At the same time, other critical exponents exhibit nontrivial behaviors distinct from the LR-GFP prediction, and the crossover at $\sigma_*=2$ may even be discontinuous, marked by a change of transition type or critical exponents~\cite{xiao_two-dimensional_2024, yao2025nonclassicalregimetwodimensionallongrange, LRHeisenberg, liu_two-dimensional_2025}. 

Analogous to the SR case, the SRW offers complementary geometric insights into the LR-SR crossover. Specifically, one considers the L\'evy flight -- a generalization of the SRW characterized by long-range jumping probabilities decaying as $p(r) \sim 1/r^{d+\sigma}$. Regarding transport properties, the LR-SRW transitions from diffusive behavior for $\sigma > 2$ to superdiffusive behavior for $0 < \sigma \le 2$. This signals a qualitative change of the connectivity of the underlying lattice at $\sigma = 2$, which is in parallel to the boundary between SR and LR universality of LR-O($n$) models and the LR-percolation model. Furthermore, the LR-SRW describes the Goldstone mode fluctuations in the long-range ordered (LRO) phase of the LR-O$(n)$ model for $n\ge2$. When the extensivity of a finite system is adequately taken into account, the LR-SRW correctly reproduces the scaling behaviors of Goldstone-mode fluctuations observed in LR-XY and LR-Heisenberg models for $0<\sigma \leq 2$ and $d=2$, as well as for any $\sigma>0$ and $d>2$~\cite{LRHeisenberg}. This geometric correspondence not only reinforces the significance of the threshold at $\sigma_*=2$ but also implies a nontrivial connection between LR-SRW mechanics and the critical phenomenon of LR models.

In parallel with these numerical breakthroughs, LR-percolation has seen significant mathematical progress~\cite{hutchcroft_dimension_2025, hutchcroft2025b, hutchcroft2025, hutchcroft2025a}. For bond percolation with connection probabilities decaying algebraically as $\sim r^{-d-\sigma}$, the asymptotics for the critical two-point function and cluster geometry have been established in various regions of the $(d,\sigma)$ plane. Notably, it has been proven that under appropriate long-range conditions, the critical two-point function behaves as $G(r) \asymp r^{-d+\sigma}$, implying $\eta = 2-\sigma$ and establishing LR mean-field behavior over a wide regime. These rigorous results serve as robust anchors for constructing the LR-percolation universality diagram and provide benchmarks for conjectures in LR-O($n$) spin and FK models, for which rigorous derivations remain lacking.

Motivated by these converging numerical, geometric, and mathematical insights, we revisit LR critical phenomena from a unified perspective. Our goal is to construct and compare universality diagrams in the $(d,\sigma)$ plane for LR-percolation, LR-O($n$) spin models, and the LR-FK–Ising model. We place particular emphasis on: (i) the structure of low-dimensional LR regimes; (ii) the onset of mean-field behavior and the hierarchy of upper critical dimensions; and (iii) the geometric properties of critical clusters, including the potential for additional geometric upper critical dimensions in the LR-FK–Ising case. Our construction synthesizes known SR and LR limits, the two-scale FSS above the upper critical dimension, insights from LR-SRW, high-precision numerical results in low dimensions~\cite{LRIsing,liu_two-dimensional_2025,yao2025nonclassicalregimetwodimensionallongrange}, and recent rigorous mathematical results~\cite{hutchcroft_dimension_2025, hutchcroft2025b, hutchcroft2025, hutchcroft2025a}. We delineate distinct regimes of critical behavior, interpreting the roles of special boundaries to propose a coherent picture connecting percolation, O($n$), and FK–Ising models.

The remainder of the paper is organized as follows. In Sec.~\ref{sec:FSS}, we summarize the finite-size scaling theory for SR critical systems, highlighting the two-scale (GFP and CG) structure and its implications for thermodynamic and geometric observables. In Sec.~\ref{sec:LR-SRW}, we describe various essential properties of LR-SRW in different regimes of $\sigma$. In Sec.~\ref{sec:LR_percolation}, we propose the universality diagram for LR-percolation in the $(d,\sigma)$ plane and discuss the associated geometric scaling. In Sec.~\ref{sec:LR_On}, we extend this analysis to LR-O($n$) spin models, discussing similarities with percolation and differences arising from internal spin symmetry. In Sec.~\ref{sec:LR_FK}, we turn to the LR-FK–Ising model, incorporating the concept of geometric upper critical dimensions to propose a corresponding universality diagram. We conclude in Sec.~\ref{sec:conclusion} with a summary and an outlook on open problems.

\section{Finite-size Scaling Theory}
\label{sec:FSS}

Finite-size scaling provides a fundamental framework for analyzing critical phenomena in numerical studies. In the thermodynamic limit, the correlation length $\xi$ diverges as $|t|^{-\nu}$, where $t=(K-K_c)/K_c$ denotes the reduced distance to criticality and $\nu$ is the correlation length exponent. This divergence gives rise to the singular behavior of both the free energy and thermodynamic observables, characterized by a set of critical exponents. In a finite system, however, the correlation length is truncated by the linear system size $L$, resulting in characteristic finite-size scaling behavior that encodes universal information about the underlying critical point. The behavior of finite-size scaling depends crucially on the spatial dimension relative to the upper critical dimension $d_c$, beyond which fluctuations become RG-irrelevant and Gaussian behavior emerges.

For systems in spatial dimensions below the upper critical dimension, the interaction is relevant in the RG sense, and nontrivial fixed points govern the universality. In this case, the singular part of the free energy density for a system obeys the standard finite-size scaling form
\begin{align}
    f(t,h,L) = L^{-d}\tilde{f}(tL^{y_t}, hL^{y_h}),
    \label{eq:ld_fss_free_energy_density}
\end{align}
where $h$ is the magnetic scaling field, $y_t \equiv 1/\nu$ and $y_h$ are the thermal and magnetic RG exponents, and $\tilde{f}(\cdot)$ is a scaling function. At criticality, the two-point correlation function exhibits the scaling behavior
\begin{align}
    g(r,L) \asymp r^{-d+2-\eta}\tilde{g}(r/L),
    \label{eq:ld_fss_correlation}
\end{align}
with $\eta = 2 +d-2 y_h$ the anomalous dimension and $\tilde{g}(\cdot)$ a scaling function. From these relations, the finite-size scaling of thermodynamic observables follows immediately. For instance, the magnetic susceptibility scales as $\chi = - \partial^2 f/\partial h^2 = L^{2y_h-d}\tilde{\chi}(tL^{y_t})$, and the specific heat capacity scales as $c_e = -\partial^2 f/\partial t^2 = L^{2y_t-d}\tilde{c}_e(tL^{y_t})$. These forms remain valid within a scaling window of width $O(L^{-y_t})$.

Above the upper critical dimension $d_c$, the GFP governs the critical behavior, and the critical exponents reduce to their mean-field values. However, the FSS in this case appears to be subtle. In particular, numerical studies of high-dimensional systems with periodic boundary conditions (PBCs) reveal a characteristic two-scaling behavior, originating from the coexistence of GFP and CG asymptotics~\cite{huang2018, fang_complete_2020, lv2021, fang_logarithmic_2021, fang_geometric_2022}. The complete graph $K_N$ consists of $N = L^d$ fully connected vertices, representing the application of Landau mean-field theory to finite systems. Above $d_c$, the CG asymptotics account for the FSS due to PBCs and influence the scaling of macroscopic observables even though the underlying field theory is Gaussian.

The free energy density above $d_c$ can then be written as
\begin{align}
    f(t,h,L) =&L^{-d}\tilde{f}_{\rm GF}(tL^{y_t^{\rm GF}}, hL^{y_h^{\rm GF}}) \notag \\ &+ L^{-d}\tilde{f}_{\rm CG}(tL^{y_t^{\rm CG}}, hL^{y_h^{\rm CG}}),
    \label{eq:hd_fss_free_energy}
\end{align}
where $(y_t^{\rm GF},y_h^{\rm GF})$ characterize the Gaussian fixed point and $(y_t^{\rm CG},y_h^{\rm CG})$ describe the FSS behavior of the complete graph~\cite{huang2018, fang_complete_2020, lv2021, fang_logarithmic_2021, fang_geometric_2022}. A similar structure appears in the two-point correlation function, which algebraically decays to a system-size-dependent plateau,
\begin{align}
    g(r,L) \asymp & \, r^{2y_h^{\rm GF}-2d}\tilde{g}_{\rm GF}(r/L) \notag \\
    &+ bL^{2y_h^{\rm CG}-2d},
    \label{eq:hd_fss_correlation}
\end{align}
with $b$ a non-universal amplitude.
The leading FSS behavior of macroscopic observables is then governed by the CG scaling, such as magnetic susceptibility $\chi = L^{2y_h^{\rm CG}-d}\tilde{\chi}(tL^{y_t^{\rm CG}})$. On the contrary, the GFP governs the FSS of distance-dependent observables. For instance, the Fourier modes of magnetic susceptibility $\chi_k \propto \sum_r e^{ikr} g(r) \sim L^{2y_h^{\rm GF}-d}\tilde{\chi}_k(tL^{y_t^{\rm GF}})$. 
These scaling forms also imply the presence of two characteristic scaling windows: a CG window of width $O(L^{-y_t^{\rm CG}})$, and a broader GFP window of width $O(L^{-y_t^{\rm GF}})$. This two-scaling structure is one of the defining features of critical phenomena above the upper critical dimension.

At the upper critical dimension itself, the Gaussian and complete-graph scaling exponents coincide, i.e., $(y_t^{\rm GF} =y_t^{\rm CG}, y_h^{\rm GF} =y_h^{\rm CG})$. The critical exponents take their mean-field values, but the finite-size scaling forms acquire multiplicative logarithmic corrections. For O($n$) models, these corrections arise from the CG contribution and lead to
\begin{align}
    f(t,h) =& L^{-d}f_{\rm GF}(tL^{y_t^{\rm GF}}, hL^{y_h^{\rm GF}}) \notag \\ 
    & + L^{-d}f_{\rm CG}(tL^{y_t^{\rm CG}}(\ln L)^{\hat{y}_t}, hL^{y_h^{\rm CG}}(\ln L)^{\hat{y}_h}),
\end{align}
where $\hat{y}_t$ and $\hat{y}_h$ are the logarithmic correction exponents~\cite{fang_logarithmic_2021,li_logarithmic_2024}. The correlation function shows an analogous structure,
\begin{align}
g(r,L) \asymp & \, r^{2y_h^{\rm GF}-2d}\tilde{g}_{\rm GF}(r/L) \notag \\
            & + bL^{2y_h^{\rm CG}-2d}(\ln L)^{\hat{y}_h}.
\end{align}
Unlike O($n$) models, where logarithmic corrections arise solely from the CG sector, for percolation, the marginality of the cubic interaction induces logarithmic factors in both the GFP and CG contributions~\cite{aharony1984, kenna2006a,kenna2006,aharony1980,essam1978, ruiz-lorenzo1998}.

In addition to thermodynamic observables, critical systems possess rich geometric properties, particularly in percolation and in the random-cluster representation of spin models. The size, shape, and spatial extent of clusters offer a complementary view of critical fluctuations and naturally lead to geometric scaling relations.
In the thermodynamic limit, the size $s$ of a critical cluster is related to its gyration radius $R$ by $s \sim R^{d_\mathrm{F}}$, where $d_\mathrm{F}$ is the fractal dimension. The distribution of cluster sizes asymptotically scales as $n(s) \sim s^{-\tau}$, with $\tau = 1+d/d_\mathrm{F}$ the Fisher exponent. For percolation, one find $d_\mathrm{F} = y_h$ when $d<6$, while for $d\ge6$, the fractal dimension is $d_\mathrm{F} = 4$ which leads to $\tau =5/2$. 

In finite systems with PBCs and $d\ge6$, the largest cluster exhibits an additional finite-size fractal dimension $d_{\mathrm{L}}$, following the complete-graph scaling,
\begin{align}
    C_1 \sim L^{d_{\mathrm{L}}} = L^{2d/3},
    \label{eq:C1_FSS}
\end{align}
If the gyration radius is measured in an unwrapped fashion, one recovers the thermodynamic relation $C_1\sim R_u^4$, where $R_u$ is the unwrapped gyration radius~\cite{fang_geometric_2022}. The scaling of the cluster-number density, which measures the distribution of cluster size in the system, is also modified by the largest cluster size as
\begin{align}
    n(s,L)\sim s^{-5/2} \tilde{n}(s/L^{2d/3}),
    \label{eq:ns_FSS}
\end{align}
where $\tilde{n}(\cdot)$ is a universal scaling function.

These geometric scaling behaviors further give rise to qualitatively different topological characteristics of critical clusters in the regimes $d<6$ and $d\ge 6$. The winding number of the largest cluster can be characterized by the ratio $R_u/L$. For $d>6$, based on $C_1 \sim R_u^4 \sim L^{2d/3}$, the unwrapped gyration radius of the largest clusters diverges as $R_u \sim L^{d/6}$. Thus, for $d>6$, the largest cluster winds around the system for an increasing number of times as $L$ grows, whereas for $d<6$ the winding number remains of order unity. Furthermore, the number of spanning clusters $N_s$, whose unwrapped gyration radius is larger than the linear system size, shows different scalings in the two cases. For $d>6$, the number of spanning clusters diverges as $N_s \propto L^d\int_{L^4}n(s, L)ds \sim L^{d-6}$~\cite{fang_geometric_2022}. However, for $d<6$, using $n(s,L)~\sim s^{-\tau}n(s/L^{d_\mathrm{F}})$ with $\tau = 1+d/d_\mathrm{F}$, one finds $N_s \sim O(1)$ in the thermodynamic limit.

The finite-size scaling behaviors reviewed above summarize how short-range systems respond to dimensionality, from standard scaling below the upper critical dimension to the two-scaling structure and logarithmic corrections that arise in higher dimensions. They also highlight general mechanisms, such as the coexistence of distinct scaling and the modification of geometric properties, that also emerge in systems with long-range couplings.

\section{Long-range Simple Random Walk}
\label{sec:LR-SRW}

While the critical behavior of interacting systems is shaped by nontrivial correlations, the transport dynamics of SRW provide a valuable heuristic baseline. This single-particle picture explicitly characterizes the GFP and reveals the effective topological connectivity of the underlying lattice. By examining how long-range steps alter the trajectory of SRW, we can gain intuitive insights into the effect of long-range interactions.


A $d$-dimensional long-range simple random walk (LR-SRW), often referred to as a L\'evy flight, is characterized by a jump probability distribution that decays algebraically with distance as $p(r) \sim r^{-(d+\sigma)}$ with $\sigma > 0$~\cite{bergersen1991,janssenLevyflightSpreadingEpidemic1999}. The transport dynamics are strictly governed by the exponent $\sigma$. Based on the convergence of the step distribution's moments, the behavior can be classified into a standard diffusive regime and a complex superdiffusive regime, which is further subdivided as follows:

\begin{itemize}
\item Diffusive Regime: $\sigma > 2$. The step distribution possesses a finite second moment, $\int p(r) r^{d-1} dr < \infty$. The central limit theorem ensures that long-range tails are irrelevant at large scales. The walk belongs to the universality class of standard Brownian motion, exhibiting normal diffusion where the mean-squared displacement (MSD) scales linearly with time or the number of steps, $\text{MSD}(t) \propto t$.

\item Superdiffusive Regime: $0 < \sigma \le 2$. When $\sigma \le 2$, the second moment diverges, and the walk is governed by L\'evy stable distributions. In this regime, the transport is characterized by the typical displacement, $R_{\text{typ}}(t) \sim t^{1/\sigma}$, where the scaling exponent $\sigma$ varies across four distinct sub-regimes:

\begin{itemize}

    \item Marginal Superdiffusive: $\sigma = 2$. At this boundary, the breakdown of the central limit theorem manifests as logarithmic corrections to the standard diffusion, leading to $R_{\text{typ}}(t) \sim (t \ln t)^{1/2}$~\cite{bouchaud1990anomalous}.
    
    \item Subballistic Regime: $1 < \sigma < 2$. Here, the first moment, i.e., the averaged step length, remains finite, but the variance diverges. The walker moves faster than diffusion but slower than a ballistic trajectory, scaling as $R_{\text{typ}}(t) \sim t^{1/\sigma}$ with $1/2 < 1/\sigma < 1$.
    
    \item Ballistic Threshold: $\sigma = 1$. This marks the point where the first moment of the step distribution begins to diverge. The dynamics become ballistic, scaling linearly as $R_{\text{typ}}(t) \sim t$. 
    
    \item Hyperballistic Regime: $0 < \sigma < 1$. The walk is dominated by extremely long-range jumps that effectively short-circuit the local lattice metric. The typical displacement grows super-linearly, $R_{\text{typ}}(t) \sim t^{1/\sigma}$ with $1/\sigma > 1$, rendering the local details of the underlying lattice negligible.
 
\end{itemize}
\end{itemize}

In summary, the diverse transport regimes of the LR-SRW provide a geometric reference point for understanding the criticality of long-range systems. The transition from diffusive to superdiffusive behavior at $\sigma=2$ is not merely a kinematic feature but reflects a topological shift in the effective connectivity of the system. Based on recent high-precision numerical results, this geometric threshold coincides with the boundary separating short-range and long-range universality classes in LR-O($n$) spin models and LR-percolation. Furthermore, the threshold at $\sigma = 1$ marks the onset of hyperballistic transport, signaling a qualitative change in the bare propagator of the LR-GFP. This divergence in the mean step length implies a more radical non-locality, potentially fundamentally altering RG flow of interactions in low-dimensional systems.

\section{Long-range percolation model}
\label{sec:LR_percolation}
\begin{figure}[ht]
  \centering
  \includegraphics[width=\linewidth]{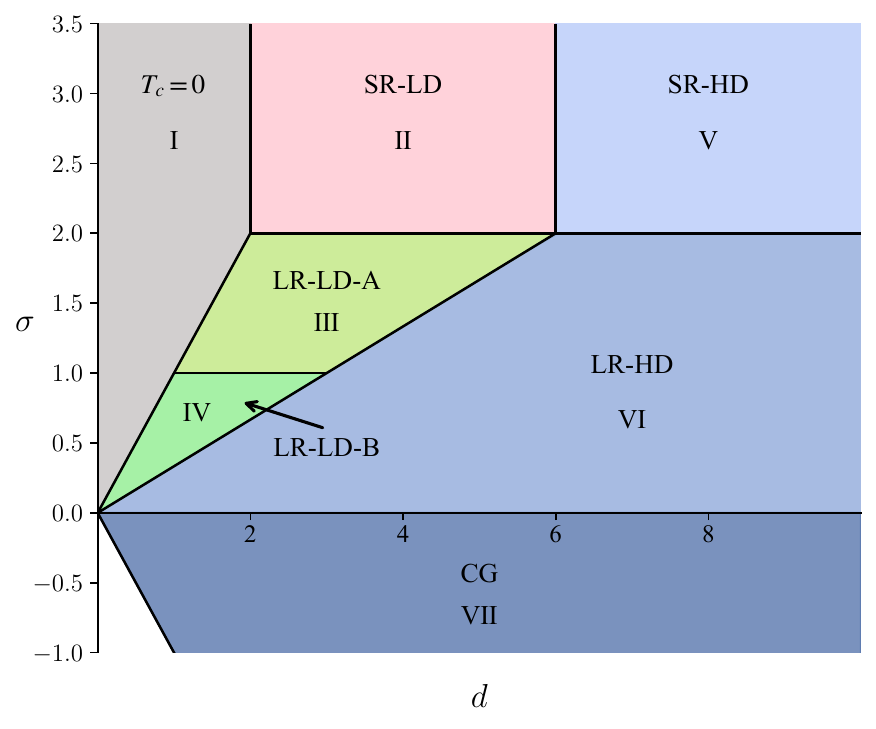}
  \caption{Universality of the long-range bond percolation in the $(d,\sigma)$ plane. Seven regimes are identified: the non-percolating region (I), the short-range universality regimes (II and V), the long-range regimes (III–VI), and the complete-graph-like region (CG, VII) with distinct scaling behaviors. The boundary at $\sigma=2$ separates the SR and LR universality classes, while $\sigma=1$ marks two different sectors in the nonclassical LR regimes.}
  \label{fig:Perco_PD}
\end{figure}

The long-range percolation model offers a natural starting point for understanding critical phenomena in systems with long-range interactions. It gives rise to vibrant thermodynamic and geometric behavior near criticality, and incorporating long-range couplings reveals an even broader spectrum of emergent collective effects. In particular, it captures geometric features across many critical systems, making it an essential reference point for exploring long-range critical physics.

We consider long-range bond percolation on a $d$-dimensional hypercubic lattice with periodic boundary conditions, where two sites $i$ and $j$ separated by $r_{ij}$ are connected with probability
\begin{align}
    p(r_{ij}) = \frac{K} { r_{ij}^{d+\sigma}},
\end{align}
with $K\in(0,1]$ controlling the bond density, analogous to the reduced coupling in O($n$) spin models~\cite{grassberger_sir_2013, grassberger2013, liu_two-dimensional_2025}. For a given pair $(d,\sigma)$, a percolation threshold $K_c(d,\sigma)$ separates a non-percolating phase from a phase in which a cluster spans the system and diverges in size in the thermodynamic limit.

In the $\sigma\to\infty$ limit, the model reduces to SR percolation, whose critical scaling behavior has been well established and is summarized in the previous section. Field-theoretically, SR percolation is governed by a $\phi^3$ theory with upper critical dimension $d_c = 6$~\cite{stauffer_introduction_2018, janssenLevyflightSpreadingEpidemic1999}. Below $d_c$, nontrivial fixed points control criticality. For $d\ge2$, the percolation transition is continuous, while for $d=1$ the transition occurs only at $K_c = 1$, corresponding to the absence of a nontrivial finite-threshold transition. By analogy with spin systems, we sometimes write $T_c = 0$ to refer to the absence of a percolation transition. Above $d_c$, the Gaussian fixed point governs the critical exponents with $y_t^{\rm GF} = 2$ and $y_h^{\rm GF} = \frac{2+d}{2}$, while periodic boundaries introduce an additional complete-graph asymptotics with $y_t^{\rm CG} = d/3$ and $y_h^{\rm CG} = 2d/3$~\cite{huang2018}. At $d = d_c$, these sectors coincide, and logarithmic corrections emerge in the finite-size scaling of both thermodynamic and geometric observables.

Introducing LR bonds can drastically change the nature of the transition. 
For $0 < \sigma \le d/3$, fluctuations are strongly suppressed and the LR Gaussian fixed point (LR-GFP) takes over. Dimensional analysis then yields mean-field-like exponents with $\nu = 1/\sigma$ and $\eta = 2 - \sigma$. For intermediate values of $\sigma$, the interplay between the $k^2$ and $k^\sigma$ terms in the propagator determines whether criticality is SR- or LR-controlled. A crossover boundary at $\sigma_* = 2 - \eta_{\rm SR}$ was predicted by field-theoretical arguments~\cite{janssenLevyflightSpreadingEpidemic1999}, whereas extensive numerical evidence in two-dimensional percolation and related O($n$) models consistently supports a sharp boundary at $\sigma_* = 2$~\cite{xiao_two-dimensional_2024, liu_two-dimensional_2025}. Complementary progress on the mathematical side shows that whenever $d > 3\sigma$ and $\sigma < 1$, the critical two-point function satisfies $G(r)\asymp r^{\sigma-d}$, implying that the relation $\eta = 2-\sigma$ is exact in this regime. More generally, this scaling is expected to hold throughout the region where an effective long-range condition remains valid. However, its extent near the SR–LR crossover, in particular for $1 < \sigma < 2$ and $d < 3\sigma$, is still undetermined~\cite{hutchcroft_dimension_2025}.

The boundary of the no-transition $(T_c = 0)$ regime can be intuitively inferred from the transport properties of LR-SRW. As discussed in Sec.~\ref{sec:LR-SRW}, the number of steps $t$ required to explore a typical length scale $R_{\rm typ}$ scales as $t \sim R_{\rm typ}^2$ for $\sigma > 2$ and $t \sim R_{\rm typ}^{\sigma}$ for $0<\sigma<2$~\cite{bouchaud1990anomalous}.
To properly address the extensivity of an $d$-dimensional finite system, we consider a configuration where the total length of the random walk scales with the system volume, i.e., $O(L^d)$. Thus, the establishment of LRO requires that the L\'evy flights can effectively span the linear system size $L$ within $O(L^d)$ steps.
For the short-range regime ($\sigma > 2$), traversing a distance $L$ requires $\sim L^2$ steps. Consequently, when $d < 2$, the total number of steps $\sim L^d$ is insufficient to support such spanning trajectories, since $L^d \ll L^2$ in the thermodynamic limit, preventing the emergence of LRO.
Conversely, in the long-range regime ($0 < \sigma < 2$), the superdiffusive transport implies that reaching distance $L$ requires only $\sim L^\sigma$ steps. For $d < \sigma$, the system lacks sufficient walk steps to bridge the distance $L$ in the thermodynamic limit, as $L^d \ll L^\sigma$. In this case, the correlations remain unable to span the system, again resulting in a disordered phase. This heuristic argument effectively delineates the boundaries of the no-transition regime at $\sigma = d$.

Collectively, these field-theoretical, numerical, and rigorous insights delineate the topology of the universality diagram shown in Fig.~\ref{fig:Perco_PD}, which contains seven regimes: $T_c=0$, SR-LD, LR-LD-A, LR-LD-B, SR-HD, LR-HD, and CG-like. Here, LD and HD denote low-dimensional and high-dimensional regimes, respectively. These regimes are separated by several structurally important boundaries that can be traced to (i) the existence of a nontrivial transition ($\sigma = d$ and $d = 2$), (ii) the SR–LR crossover ($\sigma = 2$), (iii) the relevance of interaction terms under coarse-graining ($\sigma = d/3$ and $d = 6$), and (iv) a conjectured boundary at $\sigma = 1$ that separates two types of LR regimes within the nonclassical regime.

The first two regimes are closely related to the SR percolation model in low dimensions:
\begin{itemize}
    \item Regime I ($T_c = 0$): $d < 2$ and $\sigma > d$. Based on the LR SRW argument above, we conjecture that in this regime, the correlation length remains finite and no nontrivial transition exists for finite temperature $T>0$. Namely, the criticality is anomalous and at $T_c=0$, where all the lattice sites belong to a single giant cluster. In other words, the phase transition is first-order-like, with the thermal and magnetic renormalization exponents $y_t=y_h=d$. In particular, for $d = 1$, both numerical and theoretical studies have demonstrated the absence of a phase transition for $\sigma > 1$~\cite{grassberger_sir_2013}, which is consistent with this picture.
    \item Regime II (SR-LD): $\sigma > 2$ and $2 \le d < 6$. Here, the LR tail decays sufficiently fast, and the SR interaction dominates. The phase transition is continuous and governed by conventional SR fixed points whose exponents depend only on $d$.
\end{itemize}

For $d/3 < \sigma \leq 2$ and $d < 6$, long-range interactions modify the critical scaling, resulting in exponents that vary continuously with $\sigma$, characteristic of a nonclassical LR universality sector~\cite{fisher1972}. Within this sector, numerical and rigorous insights suggest a further subdivision into two distinct regimes:
\begin{itemize}
    \item Regime III (LR-LD-A): $1 < \sigma \leq 2$ and $d/3 < \sigma < d < 6$. Here, the critical exponents vary continuously with $\sigma$. Along $\sigma = 2$ in this range, the critical behavior remains LR-dominated, producing discontinuities of exponents and universal ratios as functions of $\sigma$~\cite{liu_two-dimensional_2025}. High-precision numerical results indicate that both $y_t$ and $y_h$ deviate from the mean-field values of the LR Gaussian fixed point.
    
    \item Regime IV (LR-LD-B): $d/3 < \sigma < \min(1,d)$ for $d \in (0,3)$. Mathematical results show that $G(r)\asymp r^{-d+\sigma}$, i.e., $\eta = 2 - \sigma$, in this regime~\cite{hutchcroft_dimension_2025}, and high-precision numerical simulations confirm this prediction~\cite{LRPercolation2}. At the same time, $y_t$ remains non-mean-field-like, for which no rigorous prediction is currently available. 

    For $d=1$ and $\sigma = 1$ the transition is of Berezinskii–Kosterlitz–Thouless (BKT) type~\cite{schulman1983, grassberger_sir_2013}. 
\end{itemize}

When the long-range interaction decays sufficiently slowly, or when the dimensionality exceeds $6$, there are two additional regimes where mean-field theories govern critical scaling. In both cases, the FSS behavior is equivalent or closely analogous to the high-dimensional FSS discussed in Sec.~\ref{sec:FSS}:
\begin{itemize}
    \item Regime V (SR-HD): $\sigma > 2$ and $d \ge 6$. The SR Gaussian fixed point governs the criticality with $(y_t^{\rm GF},y_h^{\rm GF})= (2,\frac{2+d}{2})$ and $d_\mathrm{F} = 4$, but finite-size scaling is affected by the CG sector with $(y_t^{\rm CG}, y_h^{\rm CG}) = (\frac d 3, \frac {2d} 3)$. The two-scaling behavior described in Sec.~\ref{sec:FSS} is realized in this regime.
    
    \item Regime VI (LR-HD): $\sigma < \min(2, d/3)$. Criticality is determined by the LR Gaussian fixed point with $y^{\rm GF}_t=\sigma$ and $y_h^{\rm GF}=\frac{\sigma+d}{2}$, while FSS again contains a CG contribution. In particular, one expects
    \begin{align}
    \chi &\sim L^{d/3}\tilde{\chi}(tL^{d/3}), \\ 
    \chi_k &\sim L^{\sigma}\tilde{\chi}_k(tL^{\sigma}),
    \end{align}
    so that macroscopic observables are governed by the CG scaling, whereas distance-dependent quantities probe the LR-GFP scaling sector. This two-scale FSS behavior in Regime VI has been confirmed numerically~\cite{LRPercolation2}. Furthermore, rigorous results show that the fractal dimension of critical clusters in this regime is $d_{\mathrm{F}}= 2 \sigma$~\cite{hutchcroft_dimension_2025}. At the same time, the Fisher exponent remains $\tau = 5/2$ even in the complete-graph model, and we therefore expect that this value persists in the LR-HD regime. Consequently, the geometric properties are expected to obey
    \begin{align}
        C_1 &\sim L^{2d/3}, \\
        R_{u,1} &\sim L^{d/(3\sigma)},\\
        N_s &\sim L^{d-3\sigma},
    \end{align}
    where $C_1$ is the size of the largest cluster, $R_{u,1}$ its unwrapped gyration radius, and $N_s$ the number of spanning clusters.
\end{itemize}
The critical exponents that control the finite-size scaling in these six regimes are summarized in Table~\ref{tab:perco}, and the geometric critical exponents in several regimes are listed in Table~\ref{tab:perco_geo}.

Additionally, Regime VII (CG): $-d \leq \sigma \leq 0$. 
Due to the divergence of the integration $\int r^{-(d+\sigma)} r^{d-1} dr$,
one has to first consider a finite system of side length $L$ and then introduce a rescaled factor $L^{-\sigma}$ such that the bond occupation probability becomes
$p(r) = (1/L^{\sigma}) K/r^{d+\sigma}$, satisfying the extensivity of the system. 
When $\sigma = -d$, the interaction becomes distance-independent, and the lattice reduces to a complete graph (CG).
The spatial fluctuations are so suppressed
that the thermal RG exponent associated with the LR Gaussian fixed point,
$y_t^{\rm GF} = \sigma <0$, becomes irrelevant, 
and the critical behaviors are solely described by 
the CG-asymptotics, with critical exponents $(y_t^{\rm CG}, y_h^{\rm CG}) = (d/3, 2d/3)$. 

Beyond the seven primary regimes, we further draw attention to several structurally important boundaries and special points in the universality diagram, where marginality or coincidence of scaling sectors gives rise to distinctive critical corrections:
\begin{itemize}
    \item $\sigma = d$: A conjectured boundary of the no-transition region for $d \in (0,2)$, supported by Lévy-flight arguments. Along this line, the transition is conjectured to display first-order-like features.
    
    \item $(d,\sigma) = (1,1)$: The interaction is marginal and the transition is of BKT like, where the order parameter exhibits a discontinuous jump at $T_c$~\cite{grassberger_sir_2013}.
    
    \item $\sigma = 2$ for $d < 6$: Numerical studies indicate that this crossover line is governed by LR-LD fixed points, producing discontinuities in critical exponents and universal ratios as functions of $\sigma$.
    
    \item $\sigma = 2$ for $d > 6$: The SR-GFP and LR-GFP exponents coincide along this line; logarithmic corrections are expected to appear in the GFP contribution to FSS.
    
    \item $\sigma = d/3$: The marginal boundary of the LR Gaussian regime. Logarithmic corrections are anticipated, probably in both GFP and CG scalings.
 
    \item $(d,\sigma) = (6,2)$: A multicritical point where the marginal boundaries of several crossovers coincide. Logarithmic corrections are expected, potentially with forms distinct from the $\sigma = 2$ and $\sigma = d/3$ boundaries.

    \item $\sigma = 1$, for $\sigma < d < 3\sigma$: A conjectured boundary between two LR-LD regimes. Below this boundary ($\sigma \le 1$), the LR-GFP prediction of $y_h = \frac{\sigma+d}{2}$ holds, while above it, $y_h$ may deviate from this prediction.
\end{itemize}

It is worth emphasizing that the boundary $\sigma = 1$ is conjectured based on the mathematical and numerical results. Mathematically, for $d/3< \sigma \le 1$, it has been proved that $y_h = \frac{\sigma+d}{2}$ and $\eta = 2-\sigma$ strictly. However, whether this result holds in the $\sigma > 1$ regime remains an open question. High-precision Monte Carlo results show that the magnetic RG exponent is numerically consistent with the prediction $y_h = \frac{\sigma+d}{2}$ within error up to $\sigma \approx 4/3$ for $d=2$~\cite{LRPercolation2}. Taken together, these findings suggest that $\sigma = 1$ is at least a lower bound for a crossover between two qualitatively distinct LR-LD regimes. 

Intriguingly, this boundary mirrors the kinetic transition of the LR-SRW at $\sigma = 1$. As detailed in Sec.~\ref{sec:LR-SRW}, the transport dynamics shift from superdiffusive ($\sigma > 1$) to \emph{hyperballistic} ($0 < \sigma \le 1$), characterized by the typical displacement scaling $R_{\text{typ}}(t) \propto t^{1/\sigma}$. In the hyperballistic regime, the displacement grows superlinearly with time, i.e., $R_{\text{typ}}(t) \gg t$. This implies that a walker can span a spatial distance $R$ in fewer than $R$ time steps, signaling a fundamental change in the effective lattice connectivity and the bare propagator of the LR-GFP.

Collectively, these insights establish a unified universality diagram for long-range percolation, underscoring how critical behavior emerges from the interplay between lattice dimensionality and interaction range.

\begin{table}[h]
    \centering
    \renewcommand\arraystretch{1.2} 
    \caption{Summary of the critical exponents $y_t$ and $y_h$ in different regimes of the percolation model. The symbols $y_t^{\rm SR}$ and $y_h^{\rm SR}$ denote their short-range values. A check mark ($\checkmark$) indicates that the corresponding region satisfies the expected $y_t$ or $y_h$, while a cross ($\times$) indicates it does not.}
    \begin{tabular}{|l|c|c|c|c|}
    \hline
        \diagbox{Regime}{$(y_{t},\,y_{h})$} &
        \makecell[l]{~~~~~SR\\($y_t^{\rm SR},\,y_h^{\rm SR}$)} &
        \makecell[l]{~SR-GFP\\(2, $\frac{d + 2}{2}$)} &
        \makecell[l]{~LR-GFP\\($\sigma$, $\frac{d + \sigma}{2}$)} &
        \makecell[l]{~~CG\\($\frac{d}{3},\frac{2d}{3}$)} \\ \hline
        SR-LD \;\;\; (II) \hspace{0em}    & ($\checkmark$,$\checkmark$) & ($\times$,$\times$) & ($\times$,$\times$) & ($\times$,$\times$) \\ \hline
        LR-LD-A (III)   & ($\times$,$\times$) & ($\times$,$\times$) & ($\times$,$\times$) & ($\times$,$\times$) \\ \hline
        LR-LD-B (IV)   & ($\times$,$\times$) & ($\times$,$\times$) & ($\times$,$\checkmark$) & ($\times$,$\times$) \\ \hline
        SR-HD \;\;\, (V) \hspace{0em}    & ($\times$,$\times$) & ($\checkmark$,$\checkmark$) & ($\times$,$\times$) & ($\checkmark$,$\checkmark$) \\ \hline
        LR-HD \;\;\, (VI) \hspace{0em}    & ($\times$,$\times$) & ($\times$,$\times$) & ($\checkmark$,$\checkmark$) & ($\checkmark$,$\checkmark$) \\ \hline
        CG \;\;\;\;\;\;\;\;\, (VII) \hspace{0em}    & ($\times$,$\times$) & ($\times$,$\times$) & ($\times$,$\times$) & ($\checkmark$,$\checkmark$) \\
        \hline
        $T_c=0$ \;\;\;\, (I)  & \multicolumn{4}{c|}{$(y_t, y_h) = (d,d)$} \\ \hline
    \end{tabular}
    \label{tab:perco}
\end{table}

\begin{table}[h]
    \centering
    \renewcommand\arraystretch{1.2} 
    \caption{Summary of the finite-size $d_\mathrm{L}$ and thermodynamic $d_\mathrm{F}$ fractal dimensions for the largest cluster in the percolation model, where the cluster sizes and radii obey the scaling relations $C_1 \sim R_u^{d_\mathrm{F}} \sim L^{d_\mathrm{L}}$. The exponent $\tau$ indicates the Fisher exponent.}
    \begin{tabular}{|l|c|c|c|}
    \hline
        \diagbox{Regime}{Scaling} & $d_\mathrm{L}$ & $d_\mathrm{F}$ & $\tau - 1$ \\ \hline
        LR-LD-B~~(IV) & $\frac{d+\sigma}{2} $ & $\frac{d+\sigma}{2}$ & $\frac{2d}{d+\sigma}$ \\ \hline
        SR-HD~~~~~(V) & $\frac{2}{3}d$ & $4$ & $\frac{3}{2}$ \\ \hline
        LR-HD~~~~~(VI) & $\frac{2}{3}d$ & $2\sigma$ & $\frac{3}{2}$ \\ \hline
    \end{tabular}
    \label{tab:perco_geo}
\end{table}

\section{Long-range O($n$) spin model}
\label{sec:LR_On}
\begin{figure}[ht]
  \centering
  \includegraphics[width=\linewidth]{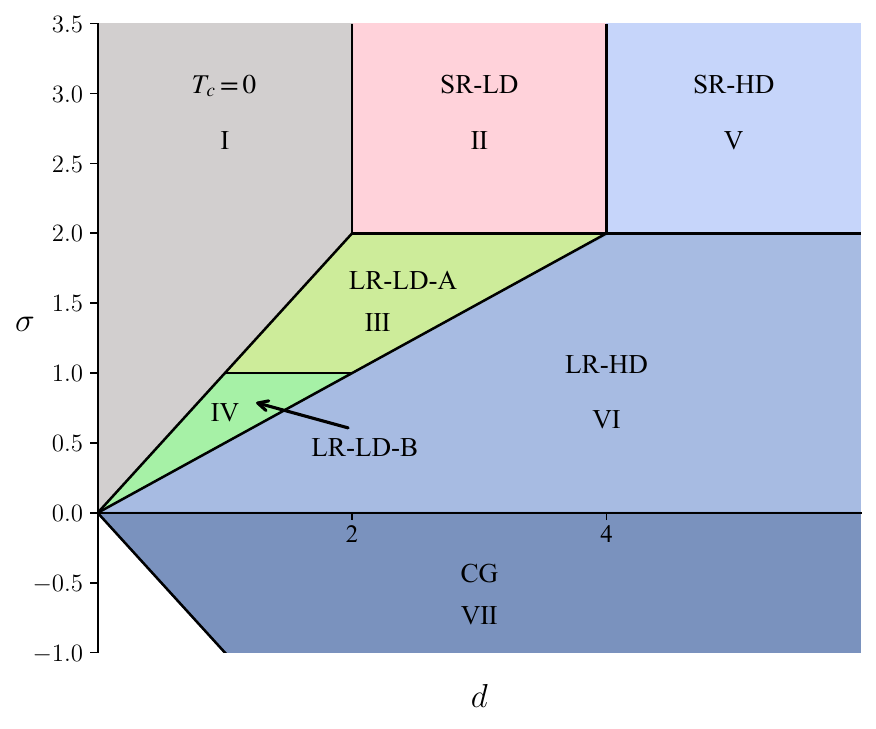}
  \caption{Universality diagram for the O($n$) models in the $(d, \sigma)$ plane. Seven regimes are identified: the no finite-temperature phase transition region (I), the short-range universality regimes (II and V), the long-range regimes (III–VI), and the complete-graph-like region (CG, VII) with distinct scaling behaviors. The boundary at $\sigma = 2$ separates the SR and LR universality classes, while $\sigma = 1$ marks a crossover between two nonclassical LR regimes.}
  \label{fig:ON_PD}
\end{figure}

We now turn to the LR-O($n$) spin models on a $d$-dimensional hypercubic lattice with periodic boundary conditions, with Hamiltonian
\begin{align}
    \mathcal{H} = - \sum_{i<j} \frac{J}{r_{ij}^{d+\sigma}} \,\mathbf{S}_i \cdot \mathbf{S}_j ,
\end{align}
where $\mathbf{S}_i$ is an $n$-component unit vector at site $i$, $N=L^d$ is the total number of sites, and $r_{ij}$ denotes the minimal-image distance on the torus~\cite{frenkel_understanding_2002,christiansen_phase_2019,agrawal_kinetics_2021}. Here, $n=1,2,3$ corresponds to the Ising, XY, and Heisenberg models, respectively. In contrast to percolation, where the order parameter is purely geometric, O($n$) models possess internal spin degrees of freedom. The competition between spatial dimension, interaction range, and internal symmetry leads to rich critical behavior.

In the SR limit $\sigma \to \infty$, the O($n$) models are governed by a $\phi^4$ field theory with upper critical dimension $d_c = 4$. For $d=1$, all O($n$) models have $T_c=0$. In $d=2$, the Mermin–Wagner theorem forbids spontaneous breaking of continuous symmetry at any finite temperature, so the nature of the transition depends on $n$: the Ising model undergoes a continuous transition into a long-range ordered phase, the XY model exhibits a BKT transition into a quasi-long-range ordered phase, and the Heisenberg model has no finite-temperature transition but shows asymptotically free behavior at low temperatures~\cite{mermin1966, kosterlitz_critical_1974,yao_asymptotic_2025}. For $d> d_c$, the SR Gaussian fixed point determines the critical scaling with exponents $y_t^{\rm GF} = 2$ and $y_h^{\rm GF} = \frac{2+d}{2}$. In finite systems with PBCs, complete-graph asymptotics with $(y_t^{\rm CG},y_h^{\rm CG})=(\frac{d}{2},\frac{3d}{4})$ are required to describe the finite-size scaling fully, as discussed in Sec.~\ref{sec:FSS}.

When long-range interactions are introduced, the nature of the transition can change substantially, in ways that are parallel to LR-percolation but modified by the internal symmetry. In $d=1$, where SR Ising, XY, and Heisenberg models have no finite-temperature transition, sufficiently slowly decaying interactions ($\sigma \le 1$) induce nontrivial criticality, and a BKT-like transition has been reported at $(d,\sigma)=(1,1)$ in the LR-Ising and LR-XY models~\cite{dysonExistencePhasetransitionOnedimensional1969, takamoto_critical_2010}. For $2 \le d < 4$, numerical and field-theoretical analyses indicate three regimes as $\sigma$ is varied. For $\sigma > 2$, the LR tail is irrelevant, and SR universality is retained. In a nonclassical regime $d/2 < \sigma \le 2$, the critical exponents become nontrivial functions of $\sigma$. Finally, for $0 < \sigma \le d/2$, the LR-GFP controls the transition, with exponents given by dimensional analysis as $(y_t^{\rm GF}, y_h^{\rm GF}) = (\sigma, (d+\sigma)/2)$. Extensive numerical evidence in $d=2$ for LR-Ising, XY, and Heisenberg systems supports a sharp SR–LR crossover at $\sigma_* = 2$~\cite{xiao_two-dimensional_2024,yao2025nonclassicalregimetwodimensionallongrange}. In particular, the point $(d,\sigma)=(2,2)$ already differs from the SR case: 2D LR-XY and LR-Heisenberg models exhibit second-order transitions, and the 2D LR-Ising exponents show detectable deviations from their SR values~\cite{xiao_two-dimensional_2024,yao2025nonclassicalregimetwodimensionallongrange, LRHeisenberg}.

Unlike the LR-percolation model, rigorous results for LR-O($n$) models in the low-dimensional nonclassical regime are lacking, but heuristic guidance can be obtained from LR-SRW. The LR-SRW effectively describes the Goldstone modes in the ordered phase of LR spin systems~\cite{LRHeisenberg}. Specifically, when incorporating a fixed number of steps of $O(L^d)$ to account for the extensivity of a finite system, the LR-SRW correctly reproduces the finite-size scaling of observables in the LRO phase, such as magnetic susceptibility. Notably, this framework succeeds even at the marginal case $\sigma=2$, where it correctly captures the existence of LRO and the logarithmic scaling of correlations that the standard Gaussian Free Field description fails to predict~\cite{xiao_two-dimensional_2024,yao2025nonclassicalregimetwodimensionallongrange, LRHeisenberg}. The LR-SRW also provides a simple geometric explanation for the no-transition regime, in analogy to the argument in the LR-percolation case. Given that the LR-SRW captures these essential physical features, the transition of the random walk from superdiffusive to hyperballistic at $\sigma=1$ leads us to conjecture a similar change in the critical behavior of O($n$) models at this boundary.

Combining SR field theory, LR-GFP analysis, numerical results, and Lévy-flight arguments yields the universality diagram shown in Fig.~\ref{fig:ON_PD}. The overall topology in the $(d,\sigma)$ plane mirrors that of LR-percolation: we again distinguish seven regimes, labeled $T_c=0$, SR-LD, LR-LD-A, LR-LD-B, SR-HD, LR-HD, and CG. The main structural differences are that the SR upper critical dimension is now $d_c=4$ (instead of $6$ for percolation) and that the boundary between LR low-dimensional and LR high-dimensional regimes is given by $\sigma = d/2$. Since several sectors closely parallel the percolation case discussed in Sec.~\ref{sec:LR_percolation}, here we only summarize the main regimes and emphasize features specific to O($n$) models.

\begin{itemize}
    \item Regime I ($T_c = 0$): $d < 2$ and $d < \sigma$. Based on LR SRW arguments, we conjecture that no finite-temperature phase transition occurs in this regime, analogous to the percolation case. In particular, for $d=1$, this is consistent with the absence of transitions for $\sigma>1$.

    \item Regime II (SR-LD): $2 < d < 4$ and $\sigma > 2$. The LR tail decays sufficiently fast that SR interactions dominate. The transition is second order for all $n$, and criticality is controlled by Wilson-Fisher fixed points with exponents $(y_t^{\rm SR},y_h^{\rm SR})$ that are independent of $\sigma$.

    \item Regime III (LR-LD-A): $\max(d/2,1) < \sigma < \min(2,d)$. LR couplings are relevant and drive a nonclassical LR universality sector in which the exponents $(y_t,y_h)$ depend on $\sigma$. The LR-GFP description no longer applies, i.e., $(y_t,y_h)\neq (\sigma,\frac{d+\sigma}2)$.

    \item Regime IV (LR-LD-B): $d/2 < \sigma < \min(1, d)$. This regime lies in the nonclassical LR sector but below the line $\sigma=1$. By analogy with the LR-percolation, and supported by LR-SRW reasoning and numerical evidence in one dimension, we conjecture that the magnetic exponent is the same as in the LR-GFP case, while the thermal exponent is not, i.e., $y_h = \frac{d+\sigma}{2}$, and $y_t \neq \sigma$.

    \item Regime V (SR-HD): $d > 4$ and $\sigma > 2$. The model lies above the SR upper critical dimension. In the thermodynamic limit, criticality is governed by the SR Gaussian fixed point with $(y_t^{\rm GF},y_h^{\rm GF}) = \left(2,\frac{d+2}{2}\right)$. For finite systems with periodic boundary conditions, an additional CG sector appears with $(y_t^{\rm CG},y_h^{\rm CG}) = \left(\frac{d}{2},\frac{3d}{4}\right)$, leading to the two-scale finite-size scaling scenario discussed in Sec.~\ref{sec:FSS}.

    \item Regime VI (LR-HD): $d>2\sigma$ and $0<\sigma<2$. The LR Gaussian fixed point with $(y^{\rm GF}_t,y^{\rm GF}_h) = \left(\sigma,\frac{d+\sigma}{2}\right)$ controls the critical scaling in the thermodynamic limit, while the CG exponents $(y_t^{\rm CG},y_h^{\rm CG})=(\frac d 2,\frac{3d}4)$ govern the finite-size scaling of global observables, as in the LR-HD regime of LR-percolation.

    \item Regime VII (CG): $-d \leq \sigma \leq 0$. The system exhibits the same 
    critical scaling behaviors as the Ising model on the complete graph, 
    governed by the exponents $(y_t^{\rm CG}, y_h^{\rm CG})=(d/2, 3d/4)$.

\end{itemize}

Beyond these bulk regimes, several structurally important boundaries and special points in the $(d,\sigma)$ plane exhibit marginal or crossover behavior. For clarity, we list them separately.

\begin{itemize}
    \item $\sigma=d$ for $d<2$: This line marks the boundary of the $T_c=0$ regime, based on LR SRW arguments. As in LR-percolation, the nature of phase transitions along this line remains unknown, in general. The point $(d,\sigma)=(1,1)$ is a notable example where the LR-Ising model exhibits BKT-like transition.
    
    \item $d=2$, $\sigma>2$: The universality reduces to the familiar 2D SR cases: Ising has a continuous transition, XY undergoes a BKT transition, and Heisenberg has no finite-temperature phase transition. At $\sigma = 2$, the nature of the transition changes: LR-XY becomes second order instead of BKT, and LR-Heisenberg acquires a finite-temperature transition.

    \item $\sigma=2$ for $2<d<4$: The line $\sigma=2$ separates SR and LR universality sectors but remains LR-dominated rather than SR-like. Numerical studies at $d=2$ show that at $(d,\sigma)=(2,2)$ the XY and Heisenberg models exhibit second-order transitions and the Ising exponents deviate from their SR values~\cite{xiao_two-dimensional_2024,yao2025nonclassicalregimetwodimensionallongrange, LRHeisenberg}. More generally, discontinuities in exponents and universal ratios as functions of $\sigma$ are expected when crossing this line within the LR-LD-A sector.

    \item $\sigma=1$ for $1\le d<2$: This boundary separates the LR-LD-A and LR-LD-B sectors and is motivated by LR SRW arguments and the analogy with LR-percolation. Its nature is still an open question. Unlike the percolation case, for O($n$) models, this boundary is accessible numerically only in $d=1$, which limits direct tests of its properties.

    \item $d=4$, $\sigma>2$: At the SR upper critical dimension, logarithmic corrections to scaling arise. In contrast to LR-percolation, these logarithmic factors primarily affect the CG scaling, while the SR-GFP contribution retains its form for $d>4$.
    
    \item $d=2\sigma$, $0<\sigma<2$: The LR analogue of the SR upper critical dimension. Here, the LR-GFP becomes marginal, and logarithmic corrections are expected, similarly to the role played by $d=4$ in the SR case.

    \item $(d,\sigma)=(4,2)$: A multicritical point where the SR–LR crossover intersects the SR upper critical dimension. Logarithmic corrections are anticipated, but their detailed form may differ from those along $d=4$ or $d=2\sigma$.
\end{itemize}

Together, these regimes and special boundaries complete the conjectured universality diagram of LR-O($n$) models and provide a natural point of comparison with the LR-percolation case discussed in Sec.~\ref{sec:LR_percolation}.

\section{Long-range FK-Ising model}
\label{sec:LR_FK}
\begin{figure}[ht]
  \centering
  \includegraphics[width=\linewidth]{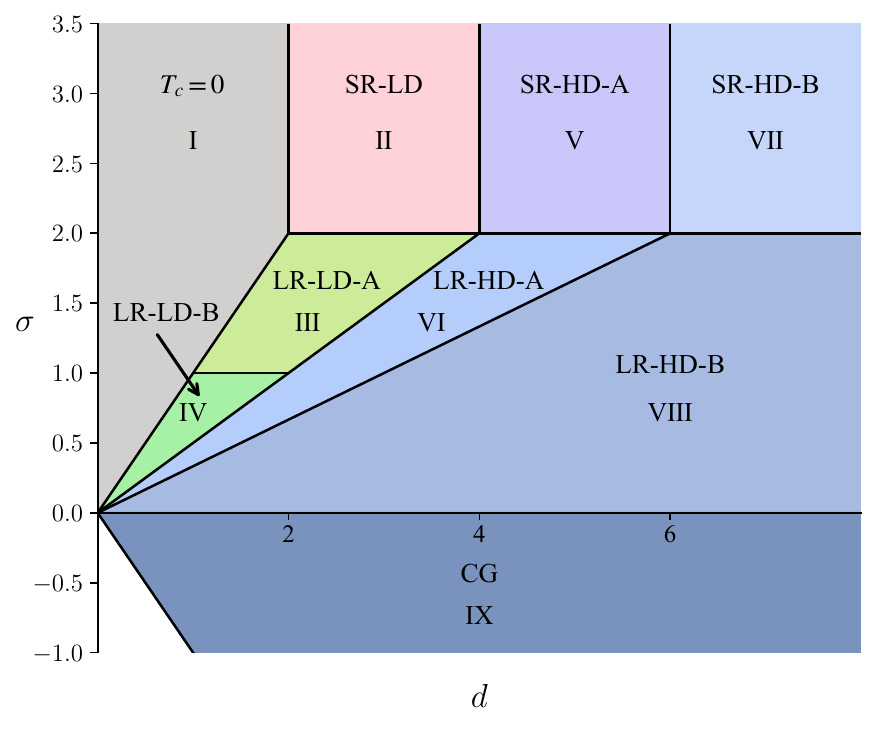}
  \caption{Universality diagram for the FK-Ising model in the $(d, \sigma)$ plane. Compared with Fig.~\ref{fig:ON_PD}, this diagram further divides the SR-HD and LR-HD regimes into SR-HD-A, SR-HD-B, LR-HD-A, and LR-HD-B, reflecting the presence of two upper critical dimensions in the FK-Ising model.}
  \label{fig:FK_PD}
\end{figure}

In addition to the standard spin representation, the Ising model admits a geometric formulation via the FK random-cluster representation~\cite{grimmett2006}, with the partition function defined in Eq.~\eqref{eq:FK}. In this framework, percolation and the Ising model share a common geometric language expressed through FK clusters.

Recent studies on the FK representation of the Ising model indicate the existence of two distinct upper critical dimensions~\cite{fang_geometric_2022, fang_geometric_2023,  wiese_two_2024, engelenburg_one-arm_2025}. The standard thermodynamic upper critical dimension, $d_c=4$, marks the onset of mean-field critical behaviors. On the other hand, the cluster geometry exhibits a second, geometric upper critical dimension at $d_p=6$, above which the critical clusters, except for the largest one, become percolation-like. Following Ref.~\cite{fang_geometric_2022}, it is convenient to parameterize the scaling of the largest and second-largest FK clusters by four fractal dimensions $(d_{\mathrm{L}1},d_{\mathrm{F}1},d_{\mathrm{L}2},d_{\mathrm{F}2})$ defined through
\begin{align}
    C_1 \sim R_1^{d_{\mathrm{F}1}} \sim L^{d_{\mathrm{L}1}}, 
    \qquad
    C_2 \sim R_2^{d_{\mathrm{F}2}} \sim L^{d_{\mathrm{L}2}},
\end{align}
where $C_1$ and $C_2$ are the sizes of the largest and second-largest FK clusters, $R_1$ and $R_2$ are their unwrapped gyration radii, and $L$ is the linear system size. Here, $d_{\mathrm{L}i}$ denotes the finite-size fractal dimension, characterizing the scaling with $L$, while $d_{\mathrm{F}i}$ represents the thermodynamic fractal dimension, relating the cluster size to its unwrapped gyration radius. For $4<d<6$, the FK clusters are geometrically Ising-like, described by $d_{\mathrm{F}1} =d_{\mathrm{L}1}=\frac{3d}{4}$ and $d_{\mathrm{F}2} =d_{\mathrm{L}2}=\frac{2+d}{2}$. The scaling of the largest cluster size is consistent with asymptotics derived rigorously on the CG Ising model~\cite{luczak_phase_2006, bollobas_random-cluster_1996}, while the scaling of the second-largest clusters is inferred from numerical results and found to be consistent with the GFP asymptotics~\cite{fang_geometric_2022, fang_geometric_2023}. The winding number of the clusters and the number of spanning clusters are both of order $O(1)$. For $d \ge 6$, the finite-size fractal dimensions $d_{\mathrm{L}1}$ and $d_{\mathrm{L}2}$ remain the same as in the lower-dimensional case, but the thermodynamic fractal dimensions of clusters become $d_{\mathrm{F}1} =9/2$ and $d_{\mathrm{F}2} =4$, independent of dimensionality. Notably, $d_{\mathrm{F}2} =4$ corresponds to high-dimensional percolation; thus, the critical clusters, except for the largest one, show percolation-like behavior.

Moreover, the cluster number density $n(s, L)$ differs in the two regimes. Generally, one expects $n(s, L)\sim s^{-\tau} \tilde{n}(s/s_{c})$, where $s_c$ is a cutoff size close to the size of the largest cluster, scaling as $s_c \sim L^{d_{\mathrm{L}1}}$. The Fisher exponent $\tau$ is believed to satisfy the hyperscaling relation $\tau = 1+ d/d_{\mathrm{L}1}$, which has been confirmed for percolation in various dimensions and for the random-cluster model for $d<4$~\cite{hou_geometric_2019}. For $4<d<6$, the Fisher exponent follows the GFP prediction as $\tau = 1+d/y_h^{\rm GF}$ with $y_h^{\rm GF} = \frac{2+d}{2}$. However, for $d\ge 6$, the hyperscaling breaks down and one finds $\tau = 5/2$, consistent with high-$d$ percolation. These differences in geometric scaling relations further lead to distinctive topological properties in the two regimes~\cite{fang_geometric_2022, fang_geometric_2023}.

The above notation naturally extends to the SR-HD regime and LR-HD regimes of the LR-Ising model. In the FK representation, the SR-HD regime of the Ising model therefore exhibits two distinct regimes in the $(d,\sigma)$ plane, denoted SR-HD-A and SR-HD-B in Fig.~\ref{fig:FK_PD}. In both regimes, the FSS behaviors of thermodynamic quantities are governed by the same set of exponents, as discussed in Sec.~\ref{sec:FSS} and Sec.~\ref{sec:LR_On}. The difference lies in the geometry of FK clusters:
\begin{itemize}
    \item Regime V (SR-HD-A): $4<d<6$ and $\sigma > 2$. In this regime, the thermodynamic and finite-size fractal dimensions are the same. The largest cluster scales with the CG-Ising exponents, while the other clusters follow GFP scaling. As a result, the unwrapped gyration radii of the largest and the second largest FK clusters scale linearly with system size, i.e., $R_1\sim R_2\sim L$. The number of spanning clusters remains bounded, $N_s=O(1)$.
    \item Regime VII (SR-HD-B): $d\ge 6$ and $\sigma > 2$. The finite-size fractal dimensions $d_{\mathrm{L}1}$ and $d_{\mathrm{L}2}$ remain the unchanged. The largest cluster has a thermodynamic fractal dimension $d_{\mathrm{F}1}=9/2$, while the second-largest cluster exhibits percolation-like behavior with $d_{\mathrm{F}2}=4$. As a result,
    \begin{align}
        R_1 \sim L^{d/6}, 
        \qquad
        R_2 \sim L^{(d+2)/8}.
    \end{align}
    The unwrapped gyration radii for both clusters grow faster than $L$, implying that these clusters wind extensively around the torus as $L$ increases. Moreover, the Fisher exponent becomes percolation-like, $\tau=5/2$, and the number of spanning clusters diverges as $N_s\sim L^{d-6}$. 
\end{itemize}

The same mechanism extends naturally to the LR-HD sector for $0 < \sigma < \min(2,d/2)$, where the FSS of thermodynamic properties is governed by the LR Gaussian fixed point with $(y_t^{\rm GF},y_h^{\rm GF})=(\sigma,(d+\sigma)/2)$ and CG-Ising asymptotics. We conjecture that there exists a $\sigma$-dependent geometric upper critical dimension at $d_p=3\sigma$ within the LR mean-field regime, which divides it into two new regimes, LR-HD-A and LR-HD-B, as shown in Fig.~\ref{fig:FK_PD}.
\begin{itemize}
    \item Regime VI (LR-HD-A): $2\sigma<d<3\sigma$ with $0 < \sigma<2$. Analogous to the SR-HD-A regime, the largest cluster follows the CG-Ising asymptotics, and the second largest clusters follow the LR-GFP scaling, i.e., $d_{\mathrm{F}1} =d_{\mathrm{L}1}=\frac{3d}{4}$ and $d_{\mathrm{F}2}=d_{\mathrm{L}2}=\frac{\sigma+d}{2}$. Both clusters scale linearly with the system size, as $R_1\sim R_2\sim L$. By analogy, we expect small clusters to have fractal dimension $d_{\mathrm{L}2}=(d+\sigma)/2$, and the Fisher exponents are $\tau = 1+2d/(d+\sigma)$. Therefore, the number of spanning clusters remains finite in the thermodynamic limit.
    
    \item Regime VIII (LR-HD-B): $d\ge 3\sigma$ and $0 < \sigma < 2$. In this regime, the critical clusters, except for the largest one, are expected to exhibit LR-percolation-like behavior, i.e., $d_{\mathrm{F}2}=2\sigma$, while the largest cluster is conjectured to have $d_{\mathrm{F}1}=\frac{9}{4}\sigma$. A notable difference from the SR-HD-B regime is that both $d_{\mathrm{F}1}$ and $d_{\mathrm{F}2}$ are $\sigma$-dependent, and the latter is consistent with those of the LR-percolation as listed in Table~\ref{tab:perco_geo}. The finite-size fractal dimensions $d_{\mathrm{L}1},d_{\mathrm{L}2}$ remain the same as in LR-HD-A. The unwrapped gyration radii of clusters are then given by,
    \begin{align}
        R_1 \sim L^{d/(3\sigma)}, 
        \qquad
        R_2 \sim L^{(d+\sigma)/(4\sigma)}.
    \end{align}
    Similar to the SR-HD-B regime, these clusters wind extensively across the periodic boundaries. Additionally, based on the insight of the CG percolation model~\cite{huang2018}, the Fisher exponent is also expected to be the same as the high-$d$ percolation case, with $\tau = 5/2$. The number of spanning clusters then grows as $N_s\sim L^{d-3\sigma}$. 
\end{itemize}

The corresponding exponents of these four regimes are collected in Table~\ref{tab:FK}. The critical properties of other regimes are identical to those of the LR-Ising model.

The line $d_p=3\sigma$ generalizes the geometric upper critical dimension to the LR regime. In the $(\sigma, d)$ plane, the FK-Ising model exhibits two upper critical dimensions: a thermodynamic one at $d_c = \min(2\sigma, 4)$, and a geometric one at $d_p = \min(3\sigma, 6)$, above which the critical clusters, except for the largest one, show percolation like scaling.

This geometric upper critical dimension can be further corroborated by the structure of scaling windows near the critical point. As detailed in Ref.~\cite{fang_geometric_2022, fang_geometric_2023} for the SR case, for $d>4$, the FK-Ising model exhibits two critical scaling windows. The first one is of size $O(L^{-d/2})$, consistent with the CG-Ising exponent $y_t^{\rm CG} = \frac{d}{2}$. In this interval, the FSS behaviors are identical to those at the critical point. However, a second scaling window exists in the high-T regime, where the FSS differs. For $4<d<6$, the size of the second window is controlled by the GFP exponent $y_t=2$. In this scaling window, the various observables follow the GFP scaling. For $d>6$, a percolation scaling window of size $O(L^{-d/3})$ emerges, where the scaling of all clusters is percolation-like. Similar scaling behaviors are expected in the LR-FK-Ising model and are summarized below.
\begin{itemize}
    \item LR-HD-A: The first scaling window is of size $O(L^{-d/2})$, while the second scaling window is controlled by the LR-GFP of size $O(L^{-\sigma})$. 
    
    \item LR-HD-B: the first scaling window is of size $O(L^{-d/2})$. The secondary window is ``percolation-like'' on the high-temperature side of the transition of size $O(L^{-d/3})$. Within this window, all clusters, including the largest one, exhibit high-$d$ percolation universality with $d_{\rm L} = 2d/3$ and $d_{\rm F} = 4$.
\end{itemize}

\begin{table}[!ht]
    \centering
    \renewcommand\arraystretch{1.4} 
    \caption{Summary of the finite-size $(d_{\mathrm{L}1}, d_{\mathrm{L}2})$ and thermodynamic $(d_{\mathrm{F}1}, d_{\mathrm{F}2})$ fractal dimensions for the largest and second-largest clusters in the FK-Ising model, where the cluster sizes and radii obey the scaling relations $C_1 \sim L^{d_{\mathrm{L}1}} \sim R_1^{d_{\mathrm{F}1}} $ and $C_2 \sim L^{d_{\mathrm{L}2}} \sim R_2^{d_{\mathrm{F}2}}$. The exponent $\tau$ indicates the Fisher exponent for small clusters.}
    \begin{tabular}{|l|c|c|c|c|c|}
    \hline
        \diagbox{Regime}{Scaling} & $d_{\mathrm{L}1}$ & $d_{\mathrm{F}1}$ & $d_{\mathrm{L}2}$ & $d_{\mathrm{F}2}$ & $\tau - 1$ \\ \hline
        SR-HD-A~~(V)     & $\frac{3}{4}d$ & $\frac{3}{4}d$ & $\frac{d+2}{2}$ & $\frac{d+2}{2}$ & $\frac{2d}{d+2}$ \\ \hline
        SR-HD-B~~(VII)   & $\frac{3}{4}d$ & $\frac{9}{2}$ & $\frac{d+2}{2}$ & $4$ & $\frac{3}{2}$ \\ \hline
        LR-HD-A~~(VI)   & $\frac{3}{4}d$ & $\frac{3}{4}d$ & $\frac{d+\sigma}{2}$ & $\frac{d+\sigma}{2}$ & $\frac{2d}{d+\sigma}$ \\ \hline
        LR-HD-B~~(VIII)     & $\frac{3}{4}d$ & $\frac{9}{4}\sigma$ & $\frac{d+\sigma}{2}$ & $2\sigma$ & $\frac{3}{2}$ \\ \hline
    \end{tabular}
    \label{tab:FK}
\end{table}

Finally, we address several key boundaries and points. First, $d=4$ ($\sigma>2$) is the SR thermodynamic upper critical dimension, where logarithmic corrections to mean-field scaling arise in thermodynamic quantities. Similar logarithmic corrections are also manifested in geometric properties; for instance, $C_1\sim L^3 (\ln L)^{1/4}$ and $C_2 \sim L^3 (\ln L)^{-1/4}$, where the latter exponent is conjectured from numerical evidence~\cite{fang_complete_2020}. Second, $d=6$ ($\sigma>2$) marks the onset of SR-HD-B scaling: FK clusters become percolation-like, and the number of spanning clusters diverges logarithmically~\cite{fang_geometric_2022}. Third, in the LR sector, the line $d=2\sigma$ plays the role of the LR thermodynamic upper critical dimension, while $d=3\sigma$ is its geometric analogue, separating LR-HD-A from LR-HD-B. Indeed, logarithmic corrections occur along the $d = 2\sigma$ line, but whether they appear along the $d = \min(3\sigma, 6)$ remains an open question.

\section{Conclusion}
\label{sec:conclusion}
In summary, we have proposed a framework for the universality of several LR classical systems. We presented comprehensive conjectured universality diagrams in the $(d, \sigma)$ plane for three canonical models: the long-range percolation model, the O($n$) spin models, and the FK-Ising model.

We integrate recent large-scale numerical simulations, rigorous mathematical results, and known field-theoretic limits (SR, GF, and CG) into a single, cohesive picture across these models. A central conjecture of our work is the division of the nonclassical LR regime (LR-LD) into two distinct phases, separated by the boundary $\sigma=1$. Based on combined evidence and insights from LR-SRW, we propose that in the regime $0 < \sigma \le 1$ (LR-LD-B), the anomalous dimension retains its LR-Gaussian prediction $\eta = 2-\sigma$, even as the thermal exponent $y_t$ deviates. 
In this regime, the hyperballistic behavior of LR-SRW implies strong nonlocality in the field-theoretical description of the model. On this basis, we conjecture that, for $0 < \sigma < 1$, the critical behaviors in the LR-percolation and LR-O($n$) models do not exhibit conformal invariance and cannot be described by conformal field theory.
In contrast, for the $1 < \sigma \le 2$ regime (LR-LD-A), we posit that all critical exponents, including $\eta$, deviate from their mean-field values.

Furthermore, by extending the known geometric properties of the SR FK-Ising model above its spin upper critical dimension ($d_c=4$) and geometric upper critical dimension $d_p=6$, we have proposed two different LR mean-field regimes. For $d/3 < \sigma < d/2$, the critical scaling is governed by CG-Ising and GF, while for $\sigma > d/3$, the scaling behaviors of critical clusters, except the largest one, are governed by LR-percolation universality.

Our perspective of long-range criticality in the $(\sigma,d)$ plane of these foundational models shall provide a conceptual framework for guiding future analytical and numerical investigations. We emphasize that, while various aspects of our picture are consistent with numerical simulations, mathematical results, and are based on physically insightful arguments, the universality diagrams in Figs.~\ref{fig:Perco_PD} -- \ref{fig:FK_PD} are conjectured and remain to be confirmed or falsified in the future. In particular, the boundaries, $\sigma = d$ for $d<2$ and $\sigma = 1$ dividing the LR-LD regime, are proposed for the first time. In addition, the boundary $\sigma=2$ contradicts Sak's criterion, but is supported by a series of numerical simulations~\cite{picco2012,xiao_two-dimensional_2024,yao2025nonclassicalregimetwodimensionallongrange, liu_two-dimensional_2025}. For the FK-Ising model in the mean-field regimes, including SR-HD-A, SR-HD-B, LD-HD-A, and LD-HD-B, a set of critical exponents has been conjectured, and the nature of logarithmic corrections at marginal boundaries remains to be investigated.

\acknowledgments
We are indebted to Timothy Garoni, Kun Chen, and Zhi-Yi Li for valuable discussions. We acknowledge the support by the National Natural Science Foundation of China (NSFC) under Grant No. 12204173 and No. 12275263, as well as the Innovation Program for Quantum Science and Technology (under Grant No. 2021ZD0301900). YD is also supported by the Natural Science Foundation of Fujian Province 802 of China (Grant No. 2023J02032).

\bibliography{ref}

\end{document}